# Engineering Delta Modeling Languages

Arne Haber
Software Engineering
RWTH Aachen University,
Germany
http://www.se-rwth.de/

Katrin Hölldobler*
Software Engineering
RWTH Aachen University,
Germany
http://www.se-rwth.de/

Carsten Kolassa
Software Engineering
RWTH Aachen University,
Germany
http://www.se-rwth.de/

Markus Look
Software Engineering
RWTH Aachen University,
Germany
http://www.se-rwth.de/

Klaus Müller
Software Engineering
RWTH Aachen University,
Germany
http://www.se-rwth.de/

Bernhard Rumpe
Software Engineering
RWTH Aachen University,
Germany
http://www.se-rwth.de/

Ina Schaefer
Software Engineering and
Automotive Informatics
TU Braunschweig, Germany
http://www.tu-bs.de/isf

## ABSTRACT

Delta modeling is a modular, yet flexible approach to capture spatial and temporal variability by explicitly representing the differences between system variants or versions. The conceptual idea of delta modeling is language-independent. But, in order to apply delta modeling for a concrete language, so far, a delta language had to be manually developed on top of the base language leading to a large variety of heterogeneous language concepts. In this paper, we present a process that allows deriving a delta language from the grammar of a given base language. Our approach relies on an automatically generated language extension that can be manually adapted to meet domain-specific needs. We illustrate our approach using delta modeling on a textual variant of statecharts.

## Categories and Subject Descriptors

D.2 [**Software**]: Software Engineering; D.2.2 [**Software Engineering**]: Design Tools and Techniques; D.2.3 [**Software Engineering**]: Coding Tools and Techniques

## 1. INTRODUCTION

Modeling is an important part of software development that allows focussing on essential system aspects in various development phases [5]. This holds for prescriptive modeling that aims at generating (parts of) software systems as well

*K. Hölldobler is supported by the DFG GK/1298 AlgoSyn.

as descriptive modeling aiming at documentation or communication issues. Modern software systems are increasingly variable to adapt to varying user requirements or environment conditions. Software product line engineering [26] is a well-established approach for developing a set of systems with commonalities and variabilities. When a product line is constructed, all modeling techniques and languages used have to support the desired variability in order to allow a seamless integration into the development process.

There are three main ways to model variability within a software product line: annotative, compositional and transformational variability modeling [34, 12]. In this paper, we focus on delta modeling, a transformational variability modeling approach [11] which contains modular, yet flexible variability modeling concepts. It can also be used to capture the evolution of software products over time [13]. In delta modeling, a set of diverse systems is represented by a designated core model and a set of deltas describing modifications to the core model. A particular product configuration is obtained by applying the changes specified in the deltas to the core model. A delta can add, remove, modify or replace elements of a model. Delta modeling has already been applied to the architecture description language MontiArc [14], to the programming language Java [29], to Class Diagrams [28], and to Simulink models [10].

Delta modeling is a generic, language-independent concept. However, to use delta-modeling in a particular development stage, the modeling and programming languages used have to be adapted to support the definition of deltas. The process of developing a delta language from a basis modeling or programming language has so far not been defined explicitly. Therefore, it depends on the knowledge and experience of the developer. For each modeling language that should be extended with a delta modeling language, very similar design steps have to be taken. The procedure itself is very time consuming, because without a streamlined derivation process the design decisions need to be made again for every delta language. With the process proposed in this paper, the delta languages are strongly related to each other

```
1  statechart Telephone {
2    initial state Idle;
3    state Active {
4      state Busy;
5      state Call;
6    }
7    Idle -> Call : [!isEngaged] numberDialed() ;
8    Idle -> Busy : [isEngaged] numberDialed() ;
9    Active -> Idle : hangUp();
10 }
```

**Listing 1: UML/P Statechart of a telephone system.**

```
1  statechart Telephone {
2    initial state Idle;
3    state Active {
4      state Voicemail;
5      state Call;
6    }
7    state Dialing;
8    Dialing -> Call : [!isEngaged] numberDialed()
         ;
9    Dialing -> Voicemail : [isEngaged]
10     numberDialed();
11   Dialing -> Voicemail : [waited5seconds]
12     numberDialed();
13   Idle -> Dialing: openLine();
14   Active -> Idle : hangUp();
15 }
```

**Listing 2: Product variant with voicemail.**

```
1  delta Voicemail {
2    modify statechart Telephone {
3      add state Dialing;
4
5      add Idle -> Dialing: openLine();
6
7      modify transition [Idle -> Call;]{
8        set source Dialing;
9      }
10
11     modify transition [Idle -> Busy;]{
12       set source Dialing;
13     }
14
15     modify state Active.Busy {
16       set name Voicemail;
17     }
18
19     add Dialing -> Voicemail:
20       [waited5seconds] numberDialed();
21   }
22 }
```

**Listing 3: Delta to add voicemail functionality.**

as they have been derived by the same process. The complexity of the delta derivation is not a problem as it is done mostly automatically within the process.

Besides the concrete syntax of a delta language, also a product variant generator has to be developed that is able to transform a core model as specified by the applied deltas.

In this paper, we introduce a process that allows to systematically derive a delta language for any textual modeling (or programming) language. The main idea is to automatically generate an initial delta language based on a common delta language by applying a set of generic derivation rules. The common delta language encapsulates the common concepts of delta modeling present in any delta language. The generated delta language can then be manually refined to meet domain-specific needs. Hence, the required effort for the manual design of a delta language is alleviated as only some adaptations have to be made. In order to demonstrate the feasibility of our approach, we realize it using the DSL toolbench MontiCore [9, 20, 21, 22]. We validate this approach by applying it to a language for modeling Statecharts and derive the corresponding delta language.

The paper is structured as follows: in Section 2, we illustrate the concept of delta modeling. Section 3 gives an overview of the language toolbench MontiCore. The process itself is described in Section 4. In Section 5, it is demonstrated by means of an example. Section 6 reviews related work and Section 7 concludes this paper.

## 2. DELTA MODELING ON STATECHARTS

First, we would like to illustrate the main concepts of delta modeling by the example of modeling variability in the UML/P [27, 30] Statecharts. The UML/P language family contains a textual representation of Statecharts similar to UML. The corresponding language for modeling UML/P Statecharts is available as a MontiCore grammar [30].

The Statechart in Listing 1 describes states and state transitions of a cellphone. The model consists of two states on the highest hierarchy level: the state `Idle` and the state `Active`. The state `Active` itself contains two inner states: the states `Busy` and `Call`. The model also contains three transitions. The first is between `Idle` and `Call` and is only traversed if the method `numberDialed()` is called and the condition `!isEngaged` evaluates to `true`. The second one is between `Idle` and `Busy` and is only traversed if the method `numberDialed()` is called and the condition `isEngaged` evaluates to `true`. The last one is between `Active` and `Idle` and is traversed when `hangUp()` is called.

In order to be able to represent variants of this statechart using deltas, we need a delta modeling language. The delta language should allow to add, remove or modify states as well as transitions and possibly other language concepts, such as preconditions, if present.

Listing 3 shows an example of this delta language applied to derive a variant of the Statechart depicted in Listing 1. The delta derives a variant of the phone that has a new state just for dialing the number and that allows recording voicemails, if the phone is already in use. In order to model that behavior, we need to define a delta (cf. l.1) which we can apply to the telephone core model. We modify the Statechart (cf. l.2) in order to add the new state `Dialing`. After that we add a new transition (cf. l.5) from the state `Idle` to the state `Dialing`. We then need to rewire two transitions by changing their source state (cf. l.7 & l.11). Additionally we modify the state `Busy` and rename it to `Voicemail` (cf. l.15). At last, we add a new transition from `Dialing` to `Voicemail` that is triggered after a certain time has elapsed (cf. l.19).

The product variant that is achieved by applying the delta `Voicemail` to our core Statechart is depicted in Listing 2. It now contains a state `Voicemail` instead of `Busy` and the added transition as well as the state `Dialing` and the corresponding transition.

As shown later, we use context conditions to check the correctness of our derived model and resolve potential un-

```
1  grammar Statechart extends Common {
2    SCDefinition =
3      "statechart" Name
4      "{" Element* "}";
5
6    interface Element;
7
8    Transition implements Element =
9      source:Name "->" target:Name
10     ( (":" TransitionBody) | ";" );
11
12   State implements Element = "state" Name ...;
13   ...
14 }
```

**Listing 4: Simplified excerpt of the UML/P Statechart grammar.**

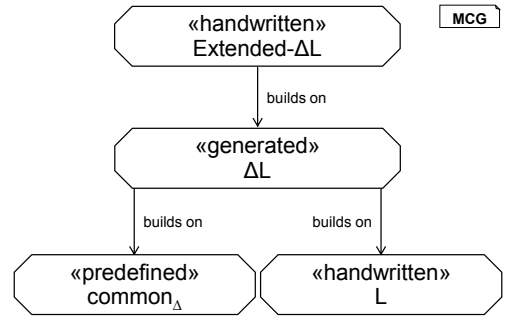

**Figure 1: Language hierarchy of concrete delta languages.**

certanties via the context.

## 3. MONTICORE LANGUAGE TOOLBENCH

The process for systematically deriving delta languages for textual modeling languages is based on the MontiCore language toolbench. We take languages defined as MontiCore grammars as input and produce delta languages that are also defined by a MontiCore grammar as result. In this section, we give a brief overview of MontiCore explaining all relevant features.

MontiCore supports the specification and generation of all relevant language processing artifacts for a specific textual language that is defined by a grammar similar to EBNF. Amongst other things, MontiCore generates the abstract and concrete syntax of a language, a lexer, a parser, and a set of runtime components, such as symbol tables and checkers for context conditions [20, 33]. A MontiCore grammar is used to define the abstract as well as the concrete syntax of a language in a single artifact.

Listing 4 shows a simplified excerpt of the Statechart grammar which defines the language that is used for modeling the UML/P [30] Statecharts in Listings 1 and 2. A MontiCore grammar starts with the keyword grammar followed by the name of the grammar (cf. l.1) and contains a set of productions defining available language elements. Listing 4 shows three productions: SCDefinition (cf. l.2), Element (cf. l.6), Transition (cf. l.8) and State (cf. l.12). SCDefinition defines the Statechart itself, State and Transition define the states and transitions of the Statechart. Element is an interface which is explained later in this section.

A production consists of a nonterminal and a right-hand side (RHS) which specifies attributes and compositions within the abstract syntax tree. As in EBNF, there might be terminals (surrounded by quotation marks (cf. l.3)) and nonterminals (cf. l.4) within the RHS. MontiCore allows to distinguish repeatedly used nonterminals by preceding the nonterminal with an identifier (cf. l.9). We also have repetition (A*,A+), alternatives (A|B), and optionality (A?).

MontiCore also facilitates language reuse by supporting modularity concepts like, e.g., language inheritance and composition (not shown here) [33, 30]. Language inheritance means that one or more existing grammars can be extended and refined by defining new grammar rules or redefining existing rules. This is denoted by the keyword extends followed by the names of the extended grammars (cf. l.1). In this way, a language developer can focus on the differences between the existing languages and the new language. To ease the reusability and extensibility of languages, it is possible to define interface-nonterminals in MontiCore grammars. An interface-nonterminal can be used like any other nonterminal within the grammar (cf. l.4) and is introduced by the keyword interface (cf. l.6). This mechanism is an extended form of alternatives. Thus the interface definition in l. 6 can be interpreted as Element = Transition | State | ..., where the RHS contains an alternative for every production that implements the interface. The language inheritance and interface concept in MontiCore is motivated by object-oriented inheritance and provides simple means to reuse and extend existing languages [19].

MontiCore also supports the definition and automatic checking of context conditions to verify that a model is well-formed. One simple context condition can, e.g., check whether the names of states within a Statechart are unique.

## 4. DERIVATION PROCESS

Based on MontiCore technology, we now introduce the process to derive a delta language for a given textual modeling language. This approach relies on the language inheritance concepts of MontiCore. Figure 1 shows the language hierarchy that is obtained when extending an existing source language L with delta modeling constructs. The basis is the abstract $common_\Delta$ language that predefines the overall structure of delta models. It additionally defines common delta operations and specifies how to identify elements in a model. The derived delta language $\Delta L$ extends this common language as well as the source language L. This way all language elements of both languages are inherited and are available in the grammar of language $\Delta L$. The automatically derived language $\Delta L$ is already complete but can also be further refined manually in order to obtain a to domain specific needs tailored delta language Extended-$\Delta L$.

### 4.1 Common Delta Constructs

The common structure for deltas is defined in the $common_\Delta$ language that we provide as a MontiCore grammar in Listing 5. The syntactical structure of a delta is defined in ll. 8 – 13. A delta has a unique name and consists of DeltaElements (cf. l.12). This interface is implemented by productions that may be used directly within a delta. Each delta has an optional ApplicationOrderConstraint (AOC) (cf. l.10). An AOC is a logical expression over delta names,

24

```
1  // Elements that may be used directly within a
2  // delta model.
3  interface DeltaElement;
4
5  // Adds concrete syntax to modifies.
6  interface ScopeIdentifier;
7
8  Delta =
9    "delta" Name
10   ("after" ApplicationOrderConstraint)?
11   "{"
12      elements:DeltaElement*
13   "}";
14
15 DeltaModify implements
16       DeltaOperation, DeltaElement =
17   "modify" ScopeIdentifier
18   modelElement:ModelElementIdentifierPath "{"
19   DeltaOperation*
20   "}";
21
22 // To identify model elements.
23 interface ModelElementIdentifier;
24
25 // Hierarchical path of MEIs.
26 ModelElementIdentifierPath =
27   parts:ModelElementIdentifier
28   ("." parts:ModelElementIdentifier)*;
29
30 // Default identifier: qualified name.
31 DefaultModelElementIdentifier implements
32   ModelElementIdentifier =
33   QualifiedModelElementName;
34
35 interface DeltaOperation;
36
37 // Operand of a delta operation.
38 interface DeltaOperand;
39
40 DeltaAdd implements DeltaOperand = "add";
41 DeltaSet implements DeltaOperand = "set";
42 DeltaRemove implements DeltaOperand = "remove";
43
44 // Default remove operation.
45 DeltaRemoveOperation implements
46       DeltaOperation =
47   DeltaRemove target:ModelElementIdentifierPath
48   ";";
```

**Listing 5:** $common_\Delta$ MontiCore grammar.

that restricts which deltas have to be applied before the current delta and which deltas must not be applied before. In the common grammar, DeltaModify (cf. ll.15–20) is the only production that implements the DeltaElement interface, therefore every Delta Element is represented by a DeltaModify. It can later also be implemented in the Extended-$\Delta$L-grammar to add further operations that may be used directly within a delta. The nonterminal ScopeIdentifier refers to an interface (cf. l.6) that is implemented by productions in the generated delta language and allows us to identify the model element which is to be modified. The nonterminal named modelElement (cf. l.18) is used to define the context that is modified by the contained DeltaOperations (cf. l.19). Modify statements defined by the production DeltaModify may contain further modify statements as this production implements the interface DeltaOperation.

A ModelElementIdentifierPath is needed to identify elements of the model. As depicted in Listing 5, it consists of dot-separated ModelElementIdentifiers named parts (cf. ll.25–28). Usually, models are hierarchically structured by a *contains* relation. Hence, the order of the parts has to reflect this hierarchical relation. Named model elements are typically identified by their name. Therefore, the default ModelElementIdentifier is a qualified name (cf. ll.30–33). Models also contain unnamed parts, e.g., transitions in a Statechart. The ModelElementIdentifier interface is implemented in a concrete delta language for each unnamed model element that has to be identified within a delta.

The interface DeltaOperation shown in Listing 5 is implemented by delta operations that may be used within a modify statement. Concrete operations must start with an operand DeltaOperand (cf. l.38) that defines the syntax of the concrete operation. Default operands are add for set-based elements of a model (cf. l.40), set for singular elements of a model (cf. l.41), and remove to delete elements from a set or to delete optional singular elements (cf. l.42). The default remove operation is given in ll. 44ff. The target of the operation is identified by a ModelElementIdentifierPath as explained above. Distinguishing between DeltaOperation and DeltaOperand allows us to generate a single production rule DeltaOperation for each nonterminal in the source language that represents all available modification operands at once.

### 4.2 Derivation Rules

Based on the source language $L$ and $common_\Delta$, we describe how to derive a delta language $\Delta L$. For new nonterminals in $\Delta L$, we use a composite name consisting of the name of the original nonterminal and the interface that is implemented, avoiding duplicate nonterminals. Within the following derivation rules, we use indices to represent this.

*Addressing Elements.*

In the delta language, it should be possible to modify every model element given by the nonterminals of the concrete language. Thus, we need to provide an implementation of the interface ModelElementIdentifier for all nonterminals $N \in L$. With the following rules, we ensure that every nonterminal can be identified, either by the default production using a qualified name or the element itself. During the automatic generation of $\Delta L$ we consider an element as addressable if it has a qualified name nonterminal with an identifier name.

1a. *For every nonterminal $N$ that can be identified by a qualified name, the default implementation of $common_\Delta$ is used to address the model element.*

1b. *For every other nonterminal $N$, the concrete syntax of the corresponding model element enclosed in brackets is used for addressing it. Thus, for $N$, we introduce a new nonterminal $\Delta N_{MEI}$ and add a production of the form:*

$$\Delta N_{MEI} \text{ implements } ModelElementIdentifier$$
$$= "[" \ N \ "]"$$

*Scope Identifier.*

The ScopeIdentifier interface of $common_\Delta$ is used to specify the element type that is addressed by the ModelElementIdentifier. With this, we are able to distinguish different model element types if they have the same

`ModelElementIdentifier` but create different scopes for the application of the delta operations. At the same time, we are able to automatically create context conditions for checking matching identifiers and types. We reuse the nonterminal of $L$ as concrete syntax of $\Delta L$. With these kind of derivation rules, we are able to formulate modify statements that identify the model element and allow modifications within this scope.

*2. For every nonterminal $N \in L$, we introduce a new nonterminal $\Delta N_{SI}$ and generate a production of the form:*

$$\Delta N_{SI} \text{ implements } \texttt{ScopeIdentifier} = "N"$$

*Delta Operation.*
With this rule, we gain the ability to specify different delta operations inside a modify statement. The abstract grammar $common_\Delta$ defines the interfaces `DeltaOperation` and `DeltaOperand`. The implementation of these interfaces is needed for every nonterminal $N$ since those are the elements that shall be modified inside a given scope.

*3. For every nonterminal $N \in L$, we introduce a new nonterminal $\Delta N_{DO}$ and generate an operation production of the form:*

$$\Delta N_{DO} \text{ implements } \texttt{DeltaOperation} = \\ \texttt{DeltaOperand } N$$

*Multiple Nonterminals.*
This derivation rule is needed since we need to consider that nonterminals might be used more than once on the RHS of a production. In MontiCore, we distinguish those by identifiers preceding the nonterminals, as shown in Section 3. For $\Delta$L we also need to distinguish these nonterminals because we would like to be able to modify them separately. We can reuse the identifiers and derive the productions for those operations. We add the nonterminal names as concrete syntax to the production to enable the distinction between the nonterminals inside the delta.

*4. For every nonterminal $N \in L$ used more than once on the RHS of a single production in $L$, we generate specific operation productions for each occurrence. For each identifier $n_i$ of $N$, we introduce a new nonterminal $\Delta n_{DO_i}$ and generate a production of the form:*

$$\Delta n_{DO_i} \text{ implements } \texttt{DeltaOperation} = \\ \texttt{DeltaOperand } "n_i" \text{ } N$$

*Delimiter Addition.*
Typically languages consist of block statements that hierarchically encapsulate other statements. Those block statements are delimited by an opening and a closing element. Inside block statements, there can be single statements that usually have a delimiter ending the statement. With a delta language, we can also modify parts of single line statements and not only complete single-line or block statements. Those parts usually have no delimiter. In this case, we add a delimiter to the corresponding delta production to achieve a uniform syntax of the delta language. For this reason, we analyze $L$ and check if the nonterminal is either a block statement or a single line statement and has, therefore, a delimiter. Otherwise, we add a final delimiter to the nonterminals in $\Delta L$.

*5. For every nonterminal $N \in L$ that is neither a block statement nor a single line statement with a line delimiter, we modify the operation production and append a delimiter.*

The derivation rules are sufficient to derive $\Delta L$ since they ensure that each nonterminal of $L$ can be addressed to be modified, and additionally, every nonterminal can be used together with an operand inside a given scope. Thus, it is possible to modify every element by adding or removing other subelements. It should be noted that the rules use concepts provided by MontiCore but are not limited to them, since the concept of interfaces can be rewritten as another production containing the alternatives, as shown in Section 3. We show the application of these rules to Statecharts in the case study in Section 5.

## 4.3 Context Conditions

In addition to the derivation rules to create the delta language $\Delta L$, we generate context conditions that provide some semantic checks for the delta language. The following enumeration provides context conditions that are automatically generated:

1. Does a `ModelElementIdentifier` reference an existing element? This can be done via resolving its qualified name and checking if there is an element with this name or via checking the complete concrete syntax of the element if used as an identifier for unnamed elements.

2. Does a `ModelElementIdentifier` reference a model element that corresponds to its type given by the `ScopeIdentifier`? This is not checked on the language level because most elements are addressed via qualified names which do not provide information about the type of the element.

3. Is a `ModelElementIdentifierPath` valid in terms of its single concatenated elements? While the previous context conditions focus on single elements, this context condition checks the path within the hierarchy of model elements.

4. Is a `DeltaOperation` applicable within the scope of its surrounding modify statement? This context condition can be inferred from the RHS of the production contained in $L$. Within the scope of the nonterminal on the left-hand side (LHS) only operations affecting nonterminals from the RHS are allowed.

5. Is a `DeltaOperand` applicable for its element? Since we use the interface `DeltaOperand` to encapsulate the available delta operations, it is possible to use either the `add` operand or the `set` operand for a model element. To ensure that `add` can only be used to add a new element to a collection and `set` can only be used for a single element, we use this context condition. It is also derivable from the production contained in $L$. Within the scope of the nonterminal used on the LHS, we can distinguish if a nonterminal on the RHS has a multiple cardinality and needs `add` as an operand, or if it has a single cardinality and thus needs `set` as an operand. This is also done for multiple nonterminals that are distinguished by given variable names.

6. Does an element that should be added not yet exist? This checking for avoidance of duplicates can be ei-

ther done via the qualified name or via equality on the attribute level of the element.

7. Does an element that should be removed exist? This check for existence of an element can be done in a similar way as the previous context condition.

### 4.4 Workflow to derive $\Delta L$

Figure 2 shows the workflow to derive a delta language and the context conditions from an existing source language as an activity diagram. First, the user provides an initial configuration in which the location of the grammar for the existing source language L is provided (**A1**). The generator reads the grammar of L (**A2**). Using the MontiCore grammar of L as an input, the generator creates the MontiCore grammar for $\Delta L$ (**A3**). This is done by adding $common_\Delta$ as super grammar (**A4**) and stepwise applying the described derivation rules explained in Subsection 4.2 to the productions of L (**A5**). After the rules have been applied we generate the context conditions for $\Delta L$ (**A6**).

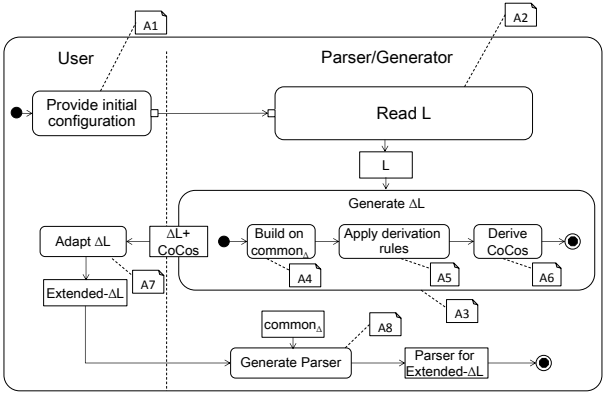

**Figure 2: Workflow of the automatic delta language derivation.**

The user can manually adapt the generated grammar $\Delta L$ by creating a subgrammar Extended-$\Delta L$ (**A7**). The manual adaption of a derived delta grammar is optional. It is useful, e.g., to tailor the syntax of the delta language. For example, $\Delta L$ may contain unneeded modify statements for language elements that should not be modifiable, or the syntax of an unnamed `ModelElementIdentifier` should be adjusted by introducing a new keyword. It is also possible to add more refined delta operations such as, a replace operation. The names of non-terminals of the source language can be part of the derived delta language. This is, for example, the case when these non-terminals need to be addressed. Sometimes the names of these non-terminals are choosen badly and need also be adapted while creating Extended-$\Delta L$. In the last step (**A8**), the base grammar $common_\Delta$ and the generated context conditions are used together with the grammar Extended-$\Delta L$ to generate the parser and the runtime components to validate the context conditions for a given delta in our new delta language.

Using the generated parser, it is now possible to parse a delta and to check that none of the context conditions has been violated.

### 4.5 Discussion

When automatically deriving a delta language $\Delta L$ from an existing source language $L$, we reuse concepts given in the language $L$. This also leads to the use of concepts of the abstract syntax of $L$ which are typically hidden from the modeler who only knows the concrete syntax. But for specifying a modify statement, the delta modeler has to know the nonterminals of $L$ as they become part of the concrete syntax of $\Delta L$. While the abstract and concrete syntax should typically be separated, we would like to present a process that automatically derives such a language $\Delta L$. Hence, the above mentioned effect cannot be completely avoided.

In order to reduce this effect, it is possible to create a hand-written language $Extended\text{-}\Delta L$ that refines $\Delta L$ and overrides the productions defining these parts of the concrete syntax. To this end, we encapsulate the concrete syntax in own productions that can be overridden. Also those productions that contain, e.g., keywords that are not suitable and should be changed can be overridden.

Reusing parts of the abstract syntax of $L$ and automatically deriving $\Delta L$ puts some requirements to the structure of $L$. For the automatic derivation of a delta language, the design of the abstract syntax is pretty important. The abstract syntax might contain folded or expanded productions that do not affect the concrete syntax of $L$. In the case of the derived $\Delta L$, the definition of productions and the use of nonterminals is important for the identification of model elements, the nesting of modify statements and the feasible delta operations within a modify block. The possible path of navigation is given by the structure of the abstract syntax and might change if the abstract syntax changes. Therefore, it might be useful to restructure the grammar of $L$, e.g. by folding or unfolding nonterminals.

We avoid nondeterminism since we use a dynamic lookahead for parsing models of the context-free grammar. In addition, the resulting grammar $\Delta L$ is always non-left-recursive by construction since the derivation rules always introduce new unique nonterminals that are only used on the LHS of the productions and never on the RHS. The new nonterminals are based on the names of the original nonterminals to prevent name clashes.

MontiCore supports language reuse by grammar extension. Thus, the source language $L$ might also be an extension from a parent language $PL$. For the derivation of the delta language from $L$, we only consider nonterminals defined directly in the language $L$ and do not consider inherited nonterminals from parent languages in this approach yet. To handle language inheritance, we assume that for every parent language $PL$ of $L$ there also exists a delta language $\Delta PL$, according to the language hierarchy, shown in Figure 3. The language $\Delta L$ builds upon $\Delta PL$ and can, therefore, also handle nonterminals defined in the super language.

## 5. CASE STUDY

In order to demonstrate our process, we use it to derive a delta language for the UML/P Statechart language introduced in Section 2. This language can then be used to describe the delta shown in Listing 3. In the following, we go through the derivation rules step by step, as described in Subsection 4.2, in order to show what each rule adds to the grammar of the delta language.

Our new delta language builds on the Statechart language and the $common_\Delta$ language. In MontiCore, this is done by extending the two languages (cf. Figure 4, part Ⓐ).

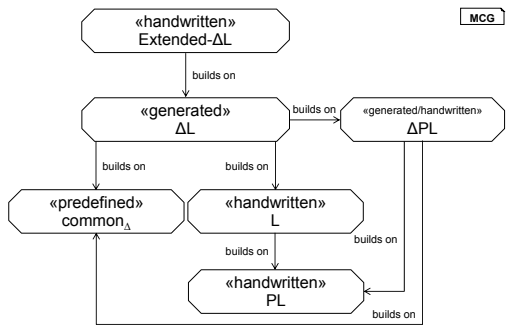

**Figure 3:** Language hierarchy of concrete delta languages extended by $PL$ and $\Delta PL$.

For the **first derivation rule**, we need to implement the `ModelElementIdentifier` interface for every nonterminal $N \in L$. In case the nonterminal $N$ has a name, we can use the default implementation in $common_\Delta$. An example for such a case is `state`, because states have names in the Statechart language. The default implementation for `ModelElementIdentifier` is shown in Listing 5. If the nonterminal $N$ has no name, which is the case for transitions, we need to introduce a new nonterminal which we call `TransitionIdentifier` that encloses the concrete syntax of a `Transition` with square brackets. The corresponding grammar is shown in Figure 4, part ①.

Using the **second derivation rule**, we derive the productions for the modify statements, which is used to denote which Statechart grammar construct we want to modify. Figure 4, part ②, shows the `ScopeIdentifiers` for `Statecharts`, `States` and `Transitions`.

With the **third derivation rule**, we specify the available operations for adding or removing transitions, see Figure 4, part ③. The delta operations are already defined in $common_\Delta$ (see Listing 5).

The original transition nonterminal consists of two name elements specifying source and target of a transition. As we would not be able to distinguish both we need to apply the **fourth derivation rule** after the third one and add the keywords "target" and "source" for the productions (see Figure 4, part ④).

The **fifth and last derivation rule** adds a delimiter to every statement that is neither a block statement nor a single line statement. In our case, this delimiter is a semicolon added by the fifth rule (cf. Figure 4, part ⑤).

However, the resulting concrete syntax of our generated delta Statechart language does not conform to the expected syntax used in Section 2 yet. For instance, the expected `ScopeIdentifier` for a transition should be "`transition`" (cf. Listing 3, ll.7 & 10) and not "`Transition`". According to the presented workflow in Subsection 4.4, we perform step (**A6**) and tailor the concrete syntax of the delta language by creating an extended delta language. This is demonstrated in Listing 6. The shown grammar builds on the generated language `DeltaStatechart` and redefines the production `DeltaTransitionScopeIdentifier`. This way, the expected syntax for the transition modification is achieved.

The grammar in Figure 4 in combination with its extension shown in Listing 6 is complete and can be used to generate a parser for our example in Listing 3.

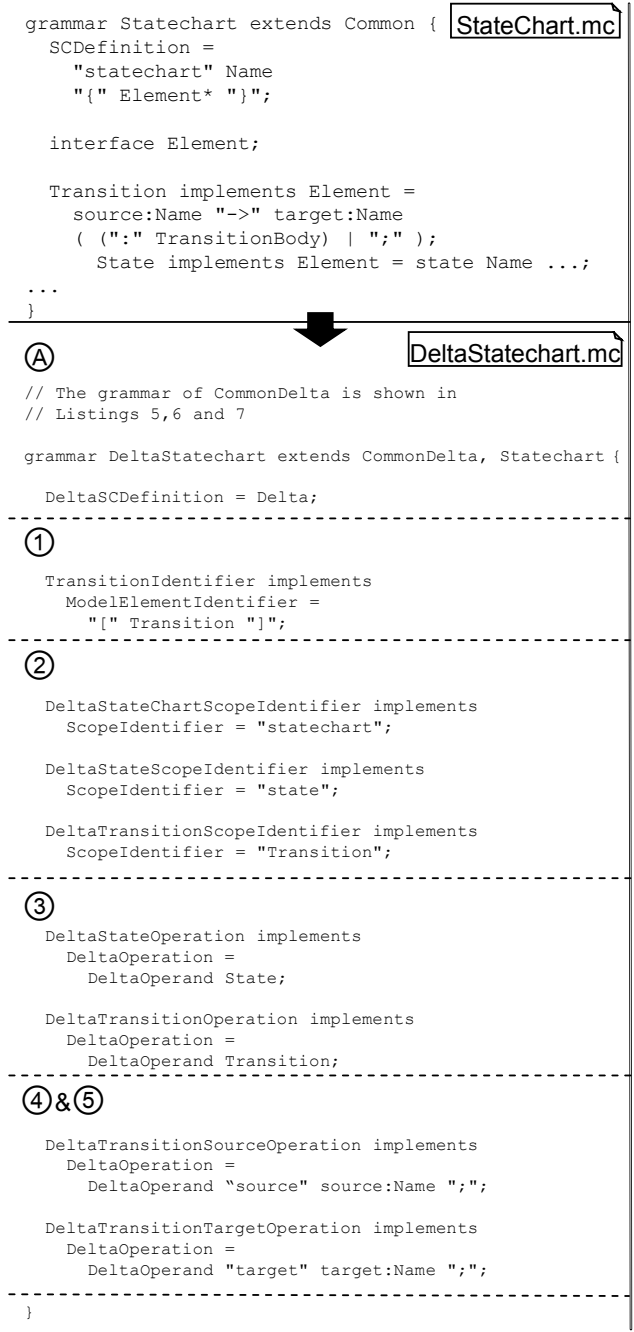

**Figure 4:** Example for the application of the derivation rules.

```
1  grammar ExtendedDeltaStatechart extends
2      DeltaStatechart {
3    DeltaTransitionScopeIdentifier implements
4        ScopeIdentifier = "transition";
5    ...
6  }
```

**Listing 6:** Extended-$\Delta L$ for the generated delta Statechart language $L$.

The first and the last line are parsed using the `Delta` production (cf. Listing 5, l.8). Within the scope of such a `Delta`, only `DeltaElements` are allowed. A modify statement is a `DeltaElement`, i.e. the statement `modify statechart Telephone ...` can be parsed using the production `DeltaModify` (cf. Listing 5, l.15) which implements `DeltaOperation` and `DeltaElement`. Additionally, `DeltaModify` requires a `ScopeIdentifier`. In our case, the `DeltaStateChartScopeIdentifier` from the $\Delta$L grammar (cf. Figure 4, part ②). Within the `Delta` production multiple `DeltaOperations` are allowed (cf. Listing 5, l.12), in our example these are the statements between line 3 and 20. They are either parsed using the `DeltaModify` production (cf. Listing 5, l.15), if they are modify statements, or by the newly generated productions `DeltaTransitionOperation` or `DeltaStateOperation`, if they are `add` statements.

As the different `add` statements are pretty similar, we just describe how one of them is parsed in detail as the other ones are parsed similarly. We use the statement from Listing 3, line 19–20, as it is the most complex one.

The whole statement is parsed using the `DeltaTransitionOperation` production (cf. Figure 4, part ③) which in turn uses the `Transition` production from the Statechart grammar and the `DeltaOperand` nonterminal which is implemented by the `DeltaAdd` nonterminal. This is illustrated in Figure 5.

The statement in line 5 is parsed exactly the same way, while the statement in line 3 is parsed similarly but using the `DeltaStateOperation` production (cf. Listing 3, ll.3 & 5). The statements that start in lines 7, 11 and 15 are modify statements (cf. Listing 3, ll.7, 11 & 15). They are parsed using the `DeltaModify` production (cf. Listing 5, l.15).

The two statements that start in lines 7 and 11 are special because they address a nonterminal that does not have a name. Therefore we cannot use the default implementation of the `ModelElementIdentifier` interface but need to generate the `TransitionIdentifier` production (cf. Figure 4, part ①) which allows us to address transitions using its complete syntax. As these statements are also parsed using the `DeltaModify` productions they can also include multiple `DeltaOperations`. In this case the `DeltaOperations` are `DeltaTransitionSourceOperations` (cf. Figure 4, part ④ & ⑤). Apart from these two differences, they are parsed similarly to the `DeltaModify` already presented.

This example shows that we can parse the delta UML/P Statechart language using our newly generated grammar. We can now parse any delta of the UML/P Statechart language. Using our process we can generate a delta language for any language with a grammar in the Monticore language toolbench.

## 6. RELATED WORK

In this section, we discuss related work in the area of variability modeling approaches, model transformation languages and methodologies, which aim at deriving transformation languages for a specific base language.

### 6.1 Variability modeling

Approaches intending to model variability in modeling languages can be classified in three main directions [34, 12]:

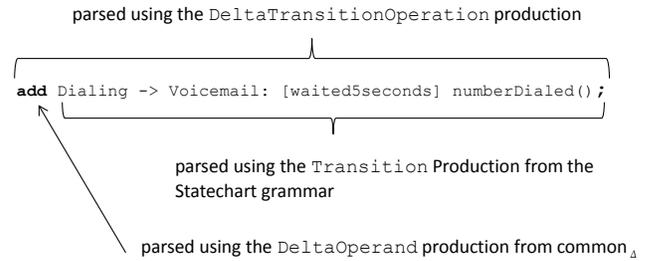

Figure 5: Example for parsing a DeltaTransitionOperation.

annotative, compositional and transformational variability modeling. Annotative approaches consider one 150% model representing all products of the product line. Variant annotations expressed using, e.g., UML stereotypes [38, 6] or presence conditions [3] define which parts of the model have to be removed to derive a concrete product model. The orthogonal variabilty model (OVM) [26] captures the variability of product line artifacts in a separate variability model in which artifact dependencies serve as annotations. A specialization of the OVM for architectural models is presented with the variability modeling language (VML) in [24].

Compositional approaches associate model fragments with product features that are composed for a particular feature configuration. In [16, 34, 25], aspect-oriented composition is used for constructing models. In [1], the composition of model fragments is performed by model superposition. In feature-oriented model-driven development [31], a combination of feature-oriented programming (FOP) and model-driven engineering (MDE), a product model is composed from a base module and a sequence of feature modules.

Transformational approaches express variability by transformation rules. The common variability language (CVF) [15] provides means to express variability of a base model in a language that does not depend on the base modeling language. This is done by specifying rules that describe how model elements of a base model have to be substituted in order to obtain a particular product model. In [18], a model composition language is introduced, which enables the specification of variant features by graph transformation rules that modify kernel models. Graph transformation rules are also used in [32, 37] to capture architectural variability. In [17], architectural variability is represented by change sets containing additions and removals of components and component connections that are applied to a base line architecture. Delta modeling also belongs to the group of transformational approaches.

Delta modeling has already been applied to several languages, like the architectural description language MontiArc [14] in [11], Java in [29], Class Diagrams in [28] and Simulink models in [10]. In contrast to these publications, our work presents a process which allows deriving a delta language from the grammar of a given base language.

### 6.2 Model transformation languages

Delta languages can be classified as a special type of model transformation languages, in which delta models correspond to transformation rules. Out of the multitude of different model transformation approaches, graph-based transformation approaches [4] are essentially the most similar to delta

29

modeling. Graph transformation rules usually consist of a LHS, a RHS and often negative application conditions (NAC). The LHS describes the pattern to be searched for in the model to be transformed and the RHS describes the pattern which replaces the matched elements. A NAC represents a pattern that must not be found. In this way, powerful transformations can be formulated.

In the following, we outline the major differences between delta languages and typical graph-based transformation languages. These differences can as well be transferred easily to other types of model transformation languages.

*(D1)* One difference concerns the need to specify NACs. In order to avoid that applying a transformation rule leads to an invalid model, the developer of a transformation rule has to specify NACs. For a transformation rule that adds a substate to a state, such a NAC can, e.g., express, that the state must not already contain a substate with the given name. Delta languages created according to our approach do not offer constructs to define NACs. This is done on purpose to simplify the specification of delta models as much as possible. Instead, we assume that these checks are implemented via context conditions that ensure that specific types of delta operations can(not) be applied. One such context condition can, e.g., ensure that adding a substate to a state is only possible if the state does not already contain a substate with the corresponding name. As presented in Subsection 4.3, some context conditions are generated and must therefore not be implemented by developers.

*(D2)* Another difference is related to the modification operations that can be specified. A delta language provides a well-defined and restricted amount of delta operations that are used for model-specific modifications. In contrast to this, graph transformation rules are capable of modeling *arbitrary* modifications. Albeit such rules are more powerful, it is easier to specify and understand the restricted amount of delta operations offered by a delta language.

*(D3)* A further difference concerns the syntax which transformation rules are based on. Most transformation languages solely operate on the abstract syntax of the models to be transformed [36]. The advantage of these approaches is that they can express transformations for *any* kind of model. However, the disadvantage is that the developer of a transformation rule inevitably needs to have a deep knowledge of the metamodel. In contrast to this, delta modeling allows reusing the concrete syntax of the corresponding modeling language.

## 6.3 Methodologies for transformation languages

In [2, 23], the metamodel of a pattern language, in which the LHS and RHS of a transformation rule are specified, is generated out of the metamodel of a modeling language. Based on this generated metamodel, the user has to define the concrete syntax for the transformation language. In [7, 8], a graphical transformation language is generated for a graphical base language. To achieve this, the user has to link the abstract syntax to the concrete syntax. In contrast to these publications, our methodology clearly defines the necessary steps to derive a grammar for a textual delta language from a textual base language. This comprises both abstract and concrete syntax of the delta language.

The most similar work to ours is the generation of a textual domain-specific transformation language (DSTL) for a textual base language described in [35]. A concrete DSTL for hierarchical automata, that could be created by this approach, is presented in [36]. In [35], the grammar for the DSTL is derived systematically from the grammar of the base language. This is comparable to our approach, that formulates how to derive a delta language systematically from the grammar of the base language. The major difference consists in the resulting languages. The differences *(D1)* and *(D2)* still hold between [35] and our approach since the applicability of transformation rules has to be restricted by NACs, and *all kinds* of model modifications and not only well-defined delta operations can be modeled. However, the difference *(D3)* does not hold as the transformation rules in [35] also reuse the concrete syntax of the corresponding modeling language.

## 7. CONCLUSION

Delta modeling is a modular, yet flexible approach to represent variability by explicitly capturing system changes. We already applied it to MontiArc [14], to Java [29], to Class Diagrams [28], and to Simulink models [10]. The general idea is language-independent. Hence, when delta modeling should be applied, for every modeling (or programming) language a separate delta language has to be designed although many of the design steps are redundant. To alleviate this, we have presented a general process to automatically derive a delta language from an existing source language. Since the MontiCore features that are used in our approach can also be within other language development frameworks [22], our approach is not limited to MontiCore, but can also be applied to other frameworks. The derived delta language can then be adapted to meet domain-specific needs. We illustrated our process by an application to the UML/P Statechart language. For future work, we aim at analyzing the requirements that have to be put on the existing language for deriving a delta language and the adaptations that have to be made to the derived language using larger case studies.